\documentclass[12pt]{article}
\usepackage[dvips]{graphicx}
\usepackage{amssymb, amsmath, mathbbol}
\usepackage[centerlast,footnotesize]{caption}

\usepackage{graphicx}
\newcommand{\One}{1\kern-4.5pt1}

\newcommand{\be}{\begin{equation}}
\newcommand{\ee}{\end{equation}}

\def\lesim{${\lower 2pt\hbox{$\scriptstyle
<$}\atop\raise 4pt\hbox{$\scriptstyle\sim$}}$} 
\def\grsim{${\lower2pt\hbox{$\scriptstyle >$} \atop\raise4pt\hbox 
{$\scriptstyle\sim$}}$} 

\begin{document}
\begin{center}
\begin{flushright}
\end{flushright}
\vskip -10mm
{\LARGE
Magnetic monopole plasma phase in (2+1)d compact quantum electrodynamics with fermionic matter
}
\vskip 1.0 cm
{{\bf  Wesley Armour$^{a,b}$, Simon Hands$^c$, John B. Kogut$^{d,e}$, Biagio Lucini$^c$, \\ 
Costas Strouthos$^f$\footnote{Corresponding author. E-mail address: strouthos@ucy.ac.cy} and 
Pavlos Vranas$^g$}}
\vskip 0.5 cm
$^a${\em Diamond Light Source, Harwell Campus, Didcot, \\
Oxfordshire OX11 0DE, United Kingdom} \\
\vskip 0.2 cm

$^b${\em Institute for the Future of Computing, Oxford Martin School, \\ 
Oxford e-Research Centre, 7 Keble Road, \\
Oxford OX1 3QG, United Kingdom} \\
\vskip 0.2 cm

$^c${\em Department of Physics, College of Science, \\ 
Swansea University, Singleton Park, Swansea SA2 8PP, United Kingdom} \\
\vskip 0.2 cm

$^d${\em Department of Energy, Division of High Energy Physics, \\
Washington,DC 20585, USA} \\

\vskip 0.2 cm
$^e${\em Department of Physics, University of Maryland, 82 Regents Drive, \\
College Park, Maryland 20742, USA} \\

\vskip 0.2 cm
$^f${\em Computation-based Science and Technology Research Center,  \\
The Cyprus Institute,
1645 Nicosia, Cyprus}

\vskip 0.2 cm
$^g${\em Lawrence Livermore National Laboratory, Livermore, CA 94550, USA}
\end{center}

\vskip 0.5 cm
\begin{center}
{\bf Abstract}
\end{center}

\noindent
We present the first evidence from lattice simulations that the magnetic
monopoles in three dimensional compact
quantum electrodynamics (cQED$_3$) with 
$N_f=2$ and $N_f= 4$ four-component fermion flavors 
are in a plasma phase.
The evidence is based mainly on the divergence of the monopole susceptibility (polarizability) 
with the lattice size at weak gauge couplings.
A weak four-Fermi term added to the cQED$_3$ action enabled
simulations with massless fermions.
The  exact chiral symmetry of the interaction terms forbids
symmetry breaking lattice discretization counter-terms to appear in the theory's effective action.
It is also shown that the scenario of a monopole plasma does not depend on the strength of the four-Fermi coupling. 
Other observables such as the densities of ``isolated'' dipoles and monopoles and the so-called specific 
heat show that a crossover from a dense monopole plasma to a dilute monopole gas occurs at 
strong couplings. The implications of our results on the stability of $U(1)$ spin liquids in 
two spatial dimensions are also discussed.
 
\newpage

\section{Introduction}
Gauge field theories play an important role in both high energy and condensed matter physics.
The mechanism of quark confinement in gauge theories with dynamical fermions such as QCD
remains one of the most elusive subjects 
in particle physics. As a result, model field theories play a significant role in 
studying this phenomenon.
Three dimensional parity-invariant compact quantum electrodynamics is such an interesting 
and challenging field theory with rich dynamics that resemble four-dimensional QCD.
It is an asymptotically free
theory, because the gauge coupling $e^2$ has mass
dimension one and thus provides the theory with a natural scale that plays the role of $\Lambda_{\rm QCD}$
in four dimensions.
Polyakov in his pioneering work on quenched cQED$_3$ \cite{Polyakov} showed analytically that 
static electric charges are confined via a 
linear potential for arbitrarily small values of the gauge coupling.
More specifically, he showed via duality 
of electric and magnetic monopole-like instanton charges that the model is equivalent to a 
three-dimensional Coulomb monopole gas described by a sine-Gordon effective action; 
this in turn leads to a
nonzero photon mass and area law for the Wilson loop \cite{Banks}.

The situation is less clear when cQED$_3$ is coupled to $N_f$ massless four-component fermionic flavors, 
because the interaction between  monopoles and antimonopoles is changed by the vacuum polarization.
A simple way of seeing why massless fermions might be expected to have a dramatic effect is to 
observe that as a result of the
Dirac quantization condition the combination $eg$ ($g$ is the charge of the magnetic monopole)
is a renormalization group (RG) invariant \cite{Calucci}. Given that the renormalized electron charge  
$e_R < e$ due to screening 
by $e^-e^+$ virtual pairs then the renormalized monopole charge $g_R > g$. 
Hence, virtual $e^-e^+$ pairs antiscreen the 
monopole-antimonopole ($m\bar{m}$) interaction. 
If the monopoles are in plasma phase at least for small $N_f$ values, then
based on the dual superconductor model \cite{'tHooft, Mandelstam}
the electric charges are linearly confined. 
In gauge field theories the particles of the vacuum that are analogous to the electrons
of the superconductor are the magnetic monopoles.
The monopoles set up magnetic currents which confine the electric field
between the charges into a narrow flux tube, in a similar way to
the electric currents around magnetic flux tubes in an ordinary superconductor. Since this
narrow flux tube has a constant energy/length, it gives rise to a linearly confining potential.

The issue of the (non-)existence of a monopole plasma phase in cQED$_3$ 
coupled to $N_f$ massless fermionic flavors has been addressed analytically by various authors.
Different approaches often based on perturbative renormalization group (RG) analysis of an  
approximate dual anomalous sine-Gordon (ASG) action led to different results depending on 
the type of approximations in the calculations.
Using an electrostatic argument and an RG calculation 
the authors of \cite{Herbut1} claimed that the interactions among
magnetic dipoles screen the logarithmic $m\bar{m}$ potential for arbitrarily large but finite $N_f$
back into the Coulomb form at large distances.
This result was confirmed by 
a self-consistent variational analysis of the dual ASG theory \cite{Herbut2}.
The results of \cite{Herbut1,Herbut2} were
criticised by the authors of \cite{Hermele2004} who showed in a systematic RG analysis that for 
large $N_f$ the monopole operators are irrelevant in the infrared limit and the physics
of the system is controlled by a conformally invariant fixed point (in the context of cuprate superconductors discussed later in 
this section it is known as
the algebraic spin liquid). 
Arguments
based on analysis of topological symmetries \cite{Unsal} produced results consistent with \cite{Hermele2004}.
In addition, the authors of \cite{Fosco} claimed that for $N_f \ge 2$ the average size of the 
$m\bar{m}$ dipoles collapses to zero leading to  non-compact QED$_3$, provided 
the fermions are massless. If the fermions have a small mass 
then the monopoles are in a dipolar phase.
In a  more recent RG calculation the authors of \cite{Nogueira} claimed that  
for $N_f > N_f^{\rm crit}=36$ the fermions are deconfined, for $20 < N_f \leq 36$
they can be either confined
or deconfined, depending on the monopole density and for $N_f \leq 20$ the fermions are 
confined. 

Lattice simulations provide a reliable non-perturbative tool for studying the role of magnetic monopoles 
in cQED$_3$. So far, there have not been any simulations that address directly
the (non-)existence of a monopole plasma phase in cQED$_3$.
The inclusion of massless fermions in the compact $U(1)$ gauge action
makes simulations difficult
due to the non-local interactions generated when integrating over fermionic variables.
Therefore, the authors of \cite{Arakawa} addressed the issue of electric charge
confinement in cQED$_3$ via lattice
simulations of an effective  $U(1)$ lattice gauge
theory with a variety of nonlocal interactions in the time-like direction
that mimic the effects of gapless/gapful
matter fields. The main result of \cite{Arakawa} is that for certain power-law decaying
interactions (mimicking coupling to massless matter fields) a second order phase transition separates
a confined from a deconfined phase. The existence of a deconfined phase in the effective
theory indicates that when cQED$_3$ is coupled to a large number of massless matter fields the theory
may be in the deconfined phase. It has been also shown with Monte Carlo simulations \cite{Kragset}
that charged particles with $\ln(r)$ interactions exhibit a phase transition at a critical temperature $T_c$
between a dilute dipole gas and a monopole plasma. 
This result also provides indirect evidence that monopoles in cQED$_3$
may be in a dipolar phase above a certain $N_f^{\rm crit}$.

In this paper we present the first attempt to resolve the controversy 
in the analytical literature
via lattice simulations of cQED$3$ with $N_f=2$ and $N_f=4$.
Massless fermion simulations
were enabled with the inclusion of a weak (unable to break chiral symmetry on its own) 
four-Fermi term in the theory's action.
The results that are largely based on the diverging monopole susceptibility (polarizability) 
with the lattice extent $L$ at weak couplings
imply that the monopoles are in a plasma phase. 
The details                                                                                   
of the lattice model including the role of the four-Fermi interaction are discussed in Sec.~2. 
Recent simulations of non-compact QED$_3$ (ncQED$_3$) with an extra weak four-fermi term \cite{Armour2010}
showed that the magnetic charges (which unlike in cQED$_3$ they are not classical solutions of the theory) 
form tightly bound $m\bar{m}$ dipoles,
because in this case
the Dirac strings carry a non-vanishing contribution to the pure gauge (non-compact) part of
the action \cite{Hands1989}. The different monopole dynamics in lattice cQED$_3$ and ncQED$_3$ for $N_f \le 4$
at weak gauge couplings imply that the two models have different continuum limits.
The authors of \cite{Fiebig, Fiore} performed simulations 
of both cQED$_3$ and ncQED$_3$
with $N_f=2$ at strong gauge couplings and concluded that the two formulations may be equivalent.
This suggestion was largely based on comparisons of chirally extrapolated data for the
chiral condensate versus monopole density which appeared to collapse on the same curve for the
two QED$_3$ formulations.
A similar claim was presented for weak couplings \cite{Fiore} based on simulations with a lattice size $L=32$. 
These results,
however, are questionable
given that $N_f=2$ ncQED$3$ simulations with $L=50$ \cite{Hands2002} 
and later on with $L=80$ \cite{Strouthos2008}
did not provide any evidence for
the existence of a nonzero chiral condensate. The principal obstruction to a definite answer in ncQED$_3$ 
is the large separation
of scales in the theory, i.e. the fermion dynamical mass is at least an order of magnitude smaller
then the natural cutoff scale $e^2$. In addition, large finite volume effects resulting from the presence of
a massless photon in the spectrum prevent a reliable extrapolation to the thermodynamic limit.
So far, the evidence from lattice simulations of ncQED$_3$ is that 
$N_f^{\rm crit} < 1.5$ \cite{Strouthos2008, Hands2004}. 
It should be noted though, that recent lattice simulations of ncQED$_3$ with an additional weak
four-Fermi term \cite{Armour2010} hinted at evidence
that chiral symmetry may be broken up to $N_f=4$.
Analytical results based on self-consistent                          
solutions of Schwinger-Dyson equations (SDE) \cite{Goecke2008, Gusynin2003} claimed 
that to detect chiral symmetry breaking for $N_f \ge 1.5$ lattice
volumes much bigger than the ones currently used in simulations
are required. The most recent SDE-based analytical calculations  
showed that $N_f^{\rm crit} \approx 4$ \cite{Goecke2008,Maris}. In addition, a gauge invariant 
calculation based on the divergence of the chiral susceptibility resulted in $N_f^{\rm crit}=2.2$ \cite{Franz}
and an RG approach on QED$_3$ with extra irrelevant four-Fermi interactions resulted 
in $N_f^{\rm crit}=6$ \cite{Kaveh}.
An alternative approach based 
on simulations of the $(2+1)d$ Thirring model at infinite coupling which may belong to the same 
universality class as ncQED$_3$ resulted in $N_f^{\rm crit}=6.6(1)$ \cite{Christofi2007}.

Although cQED$_3$ is a model field theory used for studying elementary particle physics phenomena, 
and is also an interesing basic field theory on it own right,
it acquires more concrete phenomenological significance in condensed matter physics,
because $(2+1)d$ field theories with  a compact $U(1)$ gauge field coupled to gapless relativistic
fermions arise as low energy effective field theories in
two dimensional strongly correlated electron systems, such as the cuprate superconductors 
\cite{Kim2001,Affleck1988, Lee1992, Ratner}.
Strong interactions lead to correlated electron motion resulting in unconventional states of matter
with ``fractionalized'' quantum numbers where the quasiparticle approach of Landau's Fermi liquid theory is not valid.
It is well-known from experiments that, for  cuprates, the Mott insulating state at zero doping
is the N\`{e}el antiferromagnetic state but the nature of the connection between the undoped Mott state and the
doped $d$-wave superconductor is still under theoretical debate.
In a specific incarnation of Anderson's \cite{Anderson} 
resonating valence bond idea, it was proposed \cite{Affleck1988,Lee1992}
that the so-called $U(1)$ spin liquids are 
the phases of matter that  
play a significant role in understanding underdoped cuprates. 
These featureless quantum paramagnetic states with no broken symmetries or long-range order, 
also known as critical or algebraic spin liquids (ASL),
behave as if the system is at a critical point without
the fine-tuning of any parameter. The physical picture 
can be visualized in terms of valence bonds between pairs of spin singlets separated
by arbitrarily large distances and the unpaired charge-neutral gapless spin $1/2$ spinons 
interact strongly with the valence bond
background. The flux of the emergent $U(1)$ gauge field arises from extra topological conservation 
laws not present in the microscopic theory \cite{Senthil2005}.
In a low energy description of $U(1)$ spin liquids, the spinons
with a linear dispersion are coupled minimally to an emergent
compact $U(1)$ gauge field, resulting in cQED$3$. 
Of particular importance to the physics of cuprates is the 
so-called staggered flux (or $d$-wave resonating valence bond) phase which is formally described 
by cQED$_3$ (the anisotropy of the interactions is neglected) with $N_f\!=\!2$ four-component fermions. 
Another ASL with an important role in strongly correlated electron systems is
the so-called $\pi$-flux state, which is described by $N_f\!=\!4$ cQED$_3$. 
Ghaemi and Senthil \cite{Ghaemi} showed that the staggered flux spin liquid state may be connected to the 
N\`{e}el antiferromagnetic state via a second order quantum phase transition. 
For  extensive reviews on high $T_c$
superconductivity resulting from doping a Mott insulator we refer the reader to \cite{Lee2006} 
and references therein.
These spin liquid phases are stable provided the magnetic monopoles are not in a plasma
phase and hence are unable to induce linear spinon confinement.
In addition to the works mentioned in a previous paragraph regarding the role of magnetic monopoles 
in cQED$_3$ \cite{Herbut1,Herbut2,Hermele2004, Unsal,Fosco, Nogueira}, Alicea 
\cite{Alicea} found a monopole operator which represents a symmetry allowed 
perturbation, and speculated that this may destabilize the staggered flux phase leading to charge confinement.
The numerical evidence presented in this paper in favor of the 
existence of a monopole plasma phase for $N_f \le 4$ implies that both the staggered-flux and the $\pi$-flux
spin liquids are unstable to spinon linear confinement. This in turn 
leads to  N\`{e}el antiferromagnetic order where the chiral condensate $\langle \bar{\psi}\psi \rangle$ corresponds to
the staggered magnetization.
It should also be noted that in a more  phenomenological approach an anisotropic version 
of ncQED$_3$ has been proposed 
for the phase fluctuations in the pseudogap phase (also known as algebraic Fermi liquid phase) 
of cuprate superconductors \cite{Tesanovic}.

This paper is organized as follows:
In Sec.~2 we present the lattice model, 
the various monopole observables, and the simulations parameters.
In Sec.~3 we present and discuss the simulation results  and
in Sec.~4 we present our conclusions and also
point to possible future expansions of this project.

\section{The Model}
In this first numerical exploration of the role of magnetic monopoles in cQED$_3$ with an additional 
four-Fermi term we have chosen the simplest $Z_2$ 
chirally symmetric four-Fermi interaction which for practical purposes is preferable
over terms with a continuous chiral symmetry, because the latter are not as efficiently 
simulated due to the presence of massless modes in the strongly cut-off theory. 
For computational purposes it is useful to introduce the auxiliary field
$\sigma \equiv g_s^2\bar{\chi}_i \chi_i$ (summation over the index $i$ is implied), where
$\chi_i,\bar{\chi}_i$  are $\frac{N_f}{2}$-component staggered fermion fields and $g_s^2$ is 
the four-Fermi interaction coupling.
The lattice action in terms of real-valued link potentials $\theta_{\mu\nu}$ 
is given by the following equations:
\begin{equation}
S=\sum_{i=1}^{N_f/2}\bigg(\sum_{x,x^{\prime}}\bar\chi_i(x)Q(x,x^{\prime})\chi_i(x^{\prime})\bigg)
+\frac{N_f \beta_s}{4}\sum_{\tilde x}\sigma^2_{\tilde x}
+\beta\sum_{x,\mu<\nu}(1-\cos\Theta_{\mu\nu x}),
\label{eq:action}
\end{equation}
where 
\begin{eqnarray}
\Theta_{\mu\nu x} &\equiv& \theta_\mu(x)+\theta_\nu(x+ \mu)
-\theta_\mu(x+ \nu)-\theta_\nu(x), \\
Q(x,x^\prime)&\equiv& 
\frac{1}{2}\! \sum_\mu\eta_{\mu}(x)
[e^{i\theta_{\mu }(x)}\delta_{x^\prime,x+ \mu} 
\!-\!e^{-i\theta_{\mu }(x)}\delta_{x^\prime,x- \mu} ]
+ \delta_{xx^{\prime}}\frac{1}{8}\sum_{\langle \tilde{x}, x \rangle} \sigma(\tilde{x})+m \delta_{xx^{\prime}}.
\end{eqnarray}
The indices $x$, $x^{\prime}$ consist of three integers $(x_1, x_2, x_3)$ labelling 
the lattice sites, where the third direction is considered timelike. The symbol $\langle \tilde{x}, x \rangle$ denotes the set of the eight dual lattice sites $\tilde{x}$
surrounding the direct lattice site $x$, where the $\sigma$ field lives \cite{GNM3}. The $\eta_\mu(x)$ are the Kawamoto-Smit staggered fermion phases $(-1)^{x_1+\cdots+x_{\mu-1}}$,
designed to ensure relativistic covariance of the Dirac equation in the continuum limit. 
The inverse gauge and four-Fermi couplings are given by 
$\beta \equiv \frac{1}{e^2 a}$ and $\beta_s \equiv \frac{a}{g_s^2}$, respectively, and 
$a$ is the physical lattice spacing.
The boundary conditions for the fermion fields are antiperiodic in
the timelike direction and periodic in the spatial directions. 
In the weakly coupled ($\beta \to \infty$) long-wavelength limit eq.~(\ref{eq:action}) describes $N_f$ 
four-component Dirac fermions \cite{BB}.

Performing simulations with massless fermions even with the reduced $Z_2$
chiral symmetry has substantial advantages, both theoretical and practical.
The theory has the exact symmetry of the interaction terms, which forbid chiral
symmetry breaking counterterms from appearing in its effective action.
In addition, because of the large nonzero vacuum expectation value of the
$\sigma$ field at strong gauge couplings\footnote{At strong couplings, pure QED$_3$ simulations are
 dramatically slowed down by the strong gauge field fluctuations.} or
its fluctuations at weak couplings, the Dirac operator is 
nonsingular even with $m=0$ and its inversion is very fast.
Another advantage of simulations with $m=0$, is that we do
not have to rely on often uncontrolled $m \to 0$  extrapolations to calculate  
various observables in chiral limit. 
For these reasons both the non-compact and compact lattice versions of the theory 
have been successfully simulated
in $(3+1)d$ \cite{ncqed4, cqed4} and 
showed that QED$_4$ is a logarithmically trivial theory and the systematics of the logarithms 
follow those of the Nambu$-$Jona-Lasinio model rather than those of the scalar $\lambda \phi^4$  
as often assumed. 
Unlike $(3+1)d$ where the four-Fermi term is a marginally irrelevant operator, in
$(2+1)d$ it is 
a relevant operator. It is well-known that the $(2+1)d$ Gross-Neveu model (GNM$_3$)
although non-renormalizable in the weak coupling perturbation theory, is
renormalizable in the $1/N_f$ expansion \cite{Rosenstein}. At sufficiently
strong couplings
chiral symmetry is spontaneously broken, leading to a dynamically generated fermion mass
$\Sigma \approx \langle \sigma\rangle >> m$.
The interacting continuum limit of the theory may
be taken at the critical coupling $g_{sc}^2$ (at which the gap
$\Sigma/\Lambda_{UV} \to 0$), which defines an ultraviolet-stable
renormalization group fixed point.

In QED$_3$, as the gauge coupling is varied (with the four-Fermi
coupling kept fixed at some weak  value $g_s^2<g^2_{sc}$), 
depending on the value of $N_f$ the model is  expected to undergo either
a chiral phase transition or a sharp crossover from a strong coupling
phase (where $\langle \bar{\chi} \chi \rangle \neq 0$) to a weak
coupling phase where $\langle \bar{\chi} \chi \rangle$ is either zero or
very small and possibly undetectable in lattice simulations.
Hereafter, we will use the term ``chiral transition''
to denote either a chiral phase transition or a sharp crossover from strong to weak gauge couplings.
Near the transition the weak four-Fermi term is expected to play a dominant role
as compared to the ultraviolet-finite gauge interaction.
Simulations of ncQED$_3$
\cite{Armour2010} showed that the order parameter scales with
critical exponents close to those of GNM$_3$ and the scaling region is suppressed by a factor
$\sim g_s$. 
The GNM$_3$ scaling is expected to be valid for both compact and non-compact lattice formulations. 
It should also be noted that in the large-$N_f$ and $\beta \to \infty$ limits 
the four-Fermi term is an irrelevant operator in the RG sense \cite{Kaveh}. 
 
The simulations were performed with the standard Hybrid Molecular Dynamics (HMD) R algorithm.
We used conservatively small values for the HMD trajectory time-step $dt$
and ensured that any ${\cal O}(dt^2)$ systematic
errors are smaller than the statistical errors for different observables. 
For lattice sizes smaller than 
or equal to $24^3$ we used $dt=0.005$ and an HMD trajectory length $\tau=1$ and 
for $32^3$ we used $dt=0.0025$
with a trajectory length $\tau=2$. 

The magnetic monopoles in the lattice model are identified following the standard
DeGrand and Toussaint approach \cite{DeGrand1980}.
The plaquette angles $\Theta_{\mu\nu}$ are written as
\begin{equation}
\Theta_{\mu\nu}=\bar{\Theta}_{\mu\nu}+2\pi s_{\mu\nu}(x),
\end{equation}
where $\bar{\Theta}_{\mu\nu}$ lie in the range $(-\pi,\pi]$ and $s_{\mu\nu}(x)$ is an integer that
determines the flux due to a Dirac string passing through a plaquette.
The gauge invariant integer number of monopole charges on the dual lattice sites $\tilde{x}$ are then given by
\begin{equation}
M(\tilde{x})=\epsilon_{\mu\nu\lambda}\Delta_{\mu}^+ s_{\nu \lambda}(\tilde{x}),
\end{equation}
where $\Delta_{\mu}^+$ is the forward lattice derivative and $M \in \{0,\pm 1, \pm 2\}$.
Since on a three-torus the total magnetic charge $\sum_{\tilde{x}}M(\tilde{x})=0$
we define the density of monopole charges as
\begin{equation}
\rho_M=\frac{1}{V} \sum_{\tilde{x}}|M(\tilde{x})|.
\end{equation}
The mere counting of monopoles does not provide any useful information on whether their presence 
has any impact on the  model's confining properties. As already discussed in Sec.~1  
a monopole plasma is required for linear
confinement of electric charges.
 
The observable that provides information on whether the monopoles are in a plasma or a 
dipolar phase  is the  monopole susceptibility $\chi_m$ \cite{Cardy1980}:
\begin{equation}
\label{eq:susc}
\chi_m=-\sum_{r}\langle r^2 
M(0)M(r) \rangle.
\end{equation}
The susceptibility is the polarizability of the monopole configurations; this can be readily seen 
by adding a uniform magnetic field term $-B\sum_r r M(r)$ to the dual monopole action \cite{Cardy1980} and evaluating
$\chi_m = \partial^2\ln {\cal Z}/\partial^2B|_{B=0}$, where ${\cal Z}$ is the
monopole partition function.
If the magnetic charges
are in a plasma phase, then
$\chi_m$ diverges with the lattice size $L$, implying that external magnetic fields are shielded. 
A finite $\chi_m$ means that
monopoles and antimonopoles form
a polarized gas of $m\bar{m}$ dipoles, which is what was observed in ncQED$_3$ simulations \cite{Armour2010}. 
The situation may be very different in cQED$_3$, where the monopoles are classical 
solutions of the theory and they may exist in a plasma phase at least for small $N_f$ values. Results
from numerical simulations of cQED$_3$ with $N_f=2$ and $N_f= 4$ presented in Sec.~3 favor 
the existence of a monopole plasma phase.
Also, as shown in 
\cite{Gockeler1996}, in the infinite volume limit further manipulations lead to a form of $\chi_m$ 
expressed as a Fourier transform of a two-point correlation function at zero wavevector:
\begin{equation}
\chi_m \propto \sum_x \langle \bar{\Theta}_{\mu\nu}(x)\bar{\Theta}_{\mu\nu}(0) \rangle
\propto \sum_x \langle s_{\mu\nu}(x)s_{\mu\nu}(0) \rangle.
\end{equation}  
The observable $\chi_m$  has been rarely measured in simulations with
dynamical fermions, because it is very noisy due to
near cancellations of monopole-monopole and monopole-antimonopole contributions.
With the inclusion of the four-fermi term in the QED$_3$ action the algorithm became very efficient
and $\chi_m$ has been measured with an acceptable signal-to-noise ratio at weak
gauge couplings. 
We generated $\approx 10^5 - 2\times 10^5$ configurations for the largest $L=32$ lattice 
and $\approx 3\times10^5 - 7\times 10^5$ configurations for the smaller lattices ($L=8,...,24$).

We also measured $\chi_1$ given by
\begin{equation}
\chi_1=-\langle M(0)M(1) \rangle,
\end{equation}
which includes the contributions in $\chi_m$ from adjacent lattice
cubes only. In addition, we measured $\chi_2$ given by
\begin{equation} 
\chi_2=-\sum_{r\leq \sqrt{3}}\langle r^2 M(0)M(r) \rangle,
\end{equation}
which includes the terms of $\chi_m$ where two neighboring magnetic charges share either a cube face, 
an edge or a corner.
$\chi_1 \approx \chi_2 \approx \chi_m$ indicates that the main contribution to
$\chi_m$ comes from tightly bound $m \bar{m}$ pairs. Therefore, a comparison
of $\chi_m$ with $\chi_1$ and $\chi_2$ provides information on whether the
monopole configurations are dominated by tightly bound dipoles or not. This
will become  clearer in Sec.~3 where we compare the behavior of these
observables (as a function of $L$) for both cQED$_3$ and ncQED$_3$.

\section{Results}

In this section we present results from simulations of 
the lattice model in eq.~(\ref{eq:action}) with  
$N_f=2$ and $N_f= 4$ fermion flavors. Before presenting data for the monopole observables we present results for 
the chiral condensate $\langle \bar{\chi} \chi \rangle$ versus $\beta$ near the $N_f=4$ strong coupling 
chiral transition.
In the infinite gauge coupling limit ($\beta \to 0$), it is known rigorously 
that chiral symmetry is broken \cite{Salmhofer1992} for values of $N_f$ below a certain critical value. 
Simulations of QED$_3$ with staggered fermions and $\beta=0$ showed that the theory undergoes 
a second-order phase transition at $N_f \approx 8$ \cite{Dagotto1992}. With the extra weak four-Fermi 
term the infinite gauge coupling transition is shifted towards larger $N_f$ values,
depending on the value of $\beta_s$.
Therefore, in cQED$_3$ as $\beta$ increases
and $N_f$ is larger than a putative $N_f^{\rm crit}$ (for $N_f >N_f^{\rm crit}$ the monopoles
are in the dipolar phase) there must exist a chiral 
phase transition at some critical value of the gauge coupling $\beta_c$. 
For $N_f < N_f^{\rm crit}$, since the chiral order parameter is small at weak gauge couplings, the relic
of the transition may persist as a sharp crossover between weak and strong couplings with a tail
of exponentially suppressed $\bar{\chi}\chi$ extending to weak couplings. 
The bulk of the simulations presented in this paper were performed with a 
fixed $\beta_s=2.0$, which is larger than the critical coupling $\beta_s^{\rm crit}=0.835(1)$ 
for the three-dimensional 
$N_f=4$ GNM$_3$ 
(and hence larger than $\beta_s^{\rm crit}$ for $N_f=2$ GNM$_3$).
Therefore, the four-Fermi term with $\beta_s=2.0$ cannot break chiral symmetry on its own. 
However, as already mentioned in Sec.~2, in three dimensions the weak four-Fermi term is expected 
to play a dominant role near the chiral transition as 
compared to the ultraviolet-finite gauge interaction. It was shown in simulations of ncQED$_3$ 
\cite{Armour2010} that the critical exponents extracted for the $N_f=4$ transition are close to the 
GNM$_3$ ones, and small deviations hinted at evidence for nonzero fermion mass 
generated by the gauge field dynamics.  
On a finite volume lattice near the transition the values of $\bar{\chi}\chi$ may change sign 
due to tunnelling events between the $Z_2$ vacua resulting $\langle \bar{\chi}\chi \rangle = 0$. 
In order to take into account these tunnelling events and following similar analyses of the Ising model 
 we measured the effective order 
parameter $\langle |\bar{\chi}\chi | \rangle$ instead of 
$\langle \bar{\chi}\chi  \rangle$.
We fitted the data extracted from simulations with $L=24$ 
for  $\langle |\bar{\chi}\chi | \rangle$ versus $\beta$ to the 
standard scaling relation for a second-order phase transition order parameter  
\begin{figure}[btp]
    \centering 
    \includegraphics[width=9.0cm]{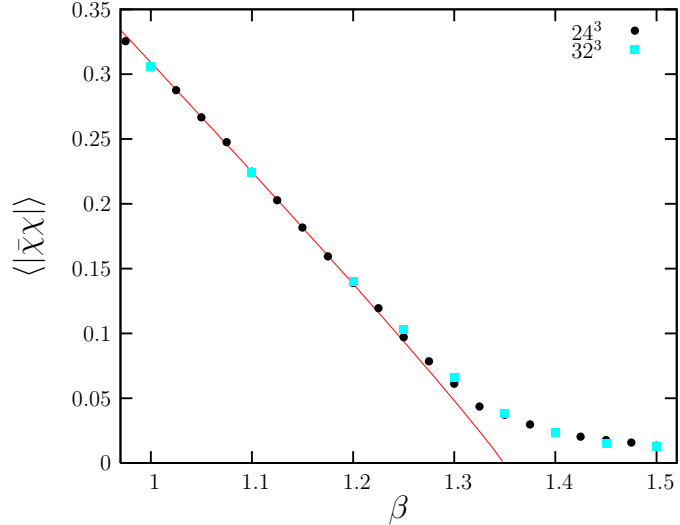}
    \caption{(Color online)
$\langle | \bar{\chi}\chi | \rangle$ vs $\beta$ extracted from simulations
with $N_f=4$ on $24^3$ and $32^3$ 
lattices. The solid curve represents the fitting function (eq.~(\ref{eq:scaling})) for $\beta \in [1.025, 1.200]$
.}
   \label{gr:condensate}
\end{figure} 

\begin{figure}[btp]
    \centering
    \includegraphics[width=9.0cm]{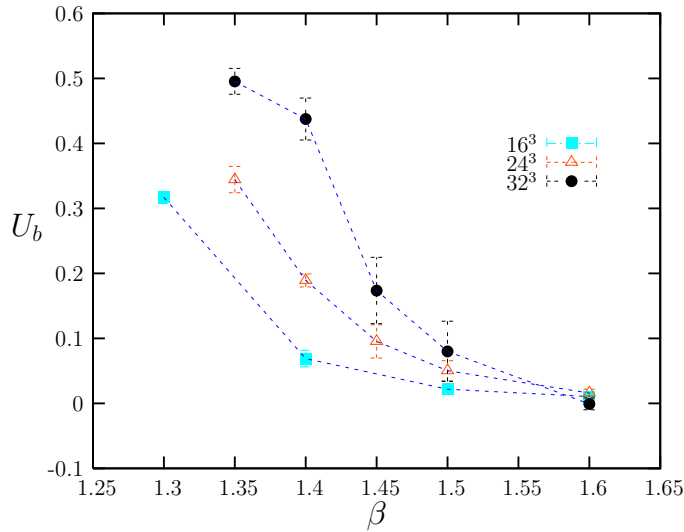}
    \caption{(Color online)
Binder's cumulant vs $\beta$ for $L=16, 24, 32$ near the $N_f=4$ chiral crossover.
}
   \label{gr:binder}
\end{figure}

\begin{table}[btp] 
\centering \caption{Values of $\langle |\bar{\chi}\chi | \rangle$ for $N_f=4$ 
from simulations on $24^3$ and $32^2$ lattices near the crossover.  
}
\medskip
\label{tab:t1}
\setlength{\tabcolsep}{1.5pc}
\begin{tabular}{ccc}
\hline \hline
$\beta$     &   $\langle |\bar{\chi}\chi | \rangle_{24}$  &  $\langle |\bar{\chi}\chi | \rangle_{32}$      \\
\hline 
1.30      &   0.0612(9)    &   0.0658(17)  \\
1.35       &  0.0372(11)  &   0.0383(15)     \\
1.40       &  0.0236(3)     &  0.0238(10)     \\
1.45       &  0.0178(5)    &  0.0154(9)       \\
1.50       &  0.0134(5)    &  0.0125(9)       \\
1.60       &  0.0112(2)    &  0.0080(2)        \\

\hline \hline
\end{tabular} 
\end{table}

\begin{equation}
\langle |\bar{\chi}\chi | \rangle = A(\beta_c-\beta)^{\beta_m}.
\label{eq:scaling}
\end{equation}
For the fitting range $\beta \in [1.025, 1.200]$ we obtained $\beta_c=1.35(1)$ and $\beta_m=0.94(3)$ with 
an acceptable fit quality given by $\chi^2/{\rm DOF}=1.8$. The extracted value of $\beta_m$ is in very good agreement
with the $\beta_m=0.93(3)$ of $N_f=4$ GNM$_3$ \cite{Christofi2007b}. The GNM$_3$ scaling confirms our earlier assertion
that the exact chiral symmetry of the lattice action forbids symmetry 
breaking lattice discretization counterterms from appearing in the model's free energy.
The $24^3$ data (together with some $32^3$ data) and the fitted curve are 
shown in Fig.~\ref{gr:condensate}.
By expanding the fitting window towards larger values of $\beta$, the value 
of $\beta_m$ increased and the fit quality deteriorated. The fit quality deteriorated dramatically 
when data points above $\beta=1.35$ were included. If, however, $\beta =1.35$ were
a critical coupling then one would expect the effective order parameter 
to obey the finite size scaling relation 
$\langle |\bar{\chi}\chi | \rangle \sim L^{\beta_m/\nu} \approx L^{-0.93}$ ($\beta_m/\nu=0.927(15)$ in $N_f=4$  GNM$_3$ \cite{Christofi2007b}).
However, as shown in Table~1, instead of observing a decrease of $\langle |\bar{\chi}\chi | \rangle$ 
with $L$ at the putative critical coupling
$\beta = 1.35$ we observe that for  $\beta=1.35, 1.40$ the values of the effective order 
parameter on $24^3$ and $32^3$ are equal within statistical errors.
The values of $\langle |\bar{\chi}\chi | \rangle$ for $\beta=1.45, 1.50$ from the two lattices
also agree within 1-2 standard deviations. Finally at $\beta=1.60$ which corresponds to the 
weakest gauge coupling in Table~1 large finite volume effects 
(the physical volume shrinks with $\beta$) result in 
$\langle |\bar{\chi}\chi | \rangle_{32} < \langle |\bar{\chi}\chi | \rangle_{24}$.
These observations together 
with the failure of eq.~(\ref{eq:scaling}) 
to provide acceptable fits when weak couplings were included in the fitting window 
constitute serious evidence that instead of 
a chiral phase transition we have a crossover from strong to weak 
couplings. It should be noted that in ncQED$_3$ although we observed small deviations of the exponents from the GNM$_3$
ones, above the transition there was a decrease of $\langle |\bar{\chi}\chi | \rangle$ with $L$, possibly because any 
tiny non-zero chiral condensate is ``swallowed'' by finite volume effects. 
This difference between the two lattice QED$_3$ formulations could 
be attributed to the different magnetic monopole dynamics, i.e. in cQED$_3$ the monopoles may be in a plasma 
phase, which in turn leads to an enhanced chiral condensate. 
Before concentrating on the role of monopoles in cQED$_3$, we will study the behavior of the  
Binder cumulant $U_b(\beta, L)$ \cite{Binder} defined by 
\begin{equation}
U_b \equiv 1 - \frac{\langle (\bar{\chi}\chi)^4 \rangle}{3 \langle (\bar{\chi}\chi)^2 \rangle^2},
\end{equation}
and measured on different lattice sizes near the chiral crossover. 
Near a second order phase transition and with sufficiently 
large lattices (where subleading corrections from finite $L$ are negligible) $U_b = f_b((\beta_c-\beta)L^{1/\nu})$. 
Therefore, at a critical coupling the lines connecting data of the same 
$L$ are expected to 
cross at a universal value $U_b=U_b^*$. In a symmetric phase as $L \to \infty$  $U_b \to 0$.  
For $N_f=4$ GNM$_3$ $U_b^*=0.232(8)$ \cite{Christofi2007}. 
In Fig.~\ref{gr:binder}  we present $U_b(\beta, L)$ data for $N_f=4$ cQED$_3$ 
with $L=16,24,32$.  
Although $U_b$ is a lot noisier than the effective order parameter it is clear that 
the lines joining data with the same $L$ 
do not cross for $\beta < 1.45$. This observation provides additional evidence in favor of 
the crossover scenario instead of a 
phase transition for $\beta < 1.5$.  The crossings of the constant $L$ $U_b$ lines occur
at $\beta \approx 1.6$ where the values of $U_b$ are less than $0.02$. 
As stated earlier 
even if for  $\beta \ge 1.6$ chiral symmetry is broken, the physical lattice volume 
is smaller and therefore 
it is plausible that finite size effects make the phase look as if it is symmetric.
\begin{figure}[tb]
\begin{center}
\begin{minipage}[c][7.8cm][c]{7.8cm}
\begin{center}
    \includegraphics[width=7.8cm]{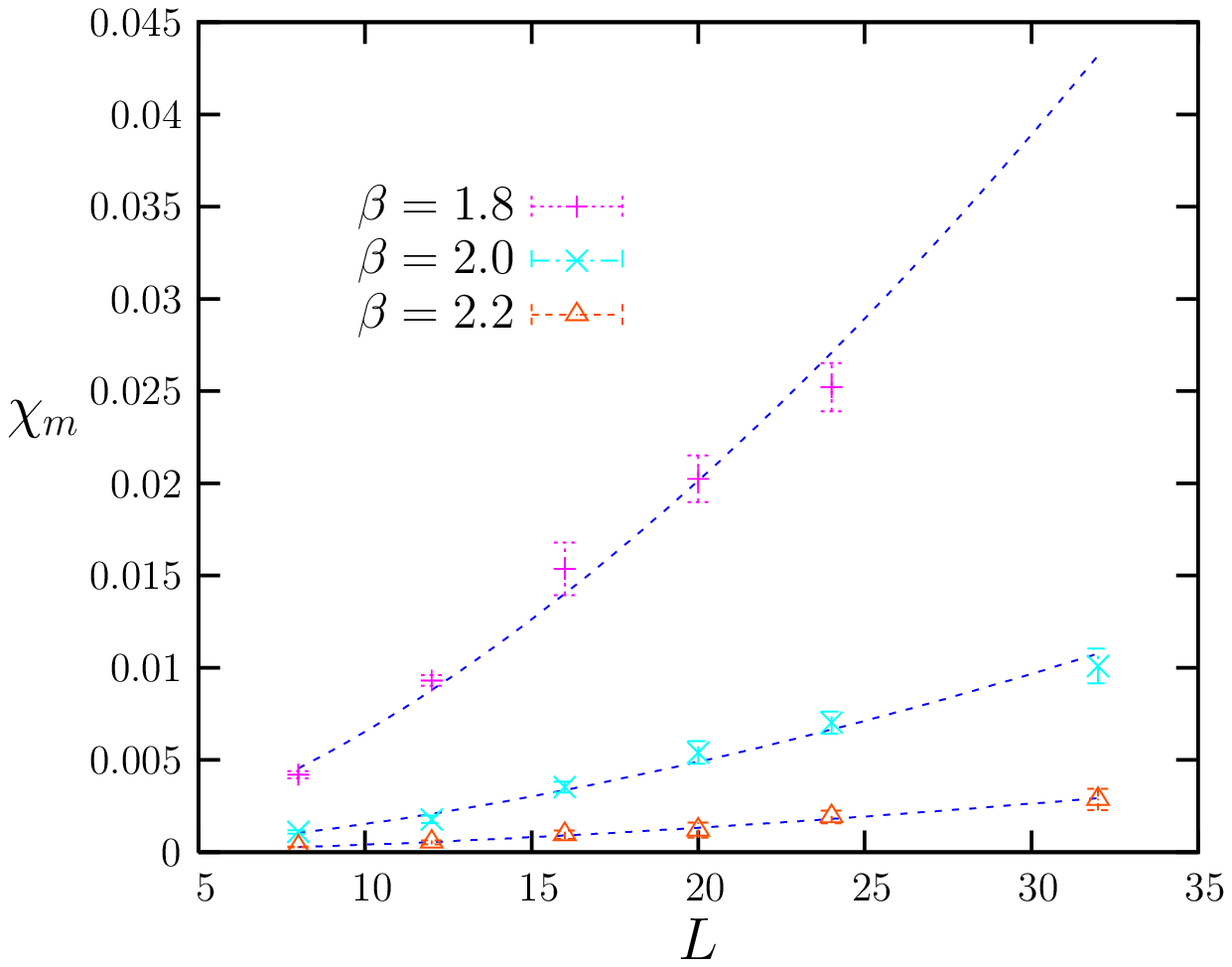}
\\[-1mm]
(a) 
\end{center}
\end{minipage}
\begin{minipage}[c][7.8cm][c]{7.8cm}
\begin{center}
    \includegraphics[width=7.8cm]{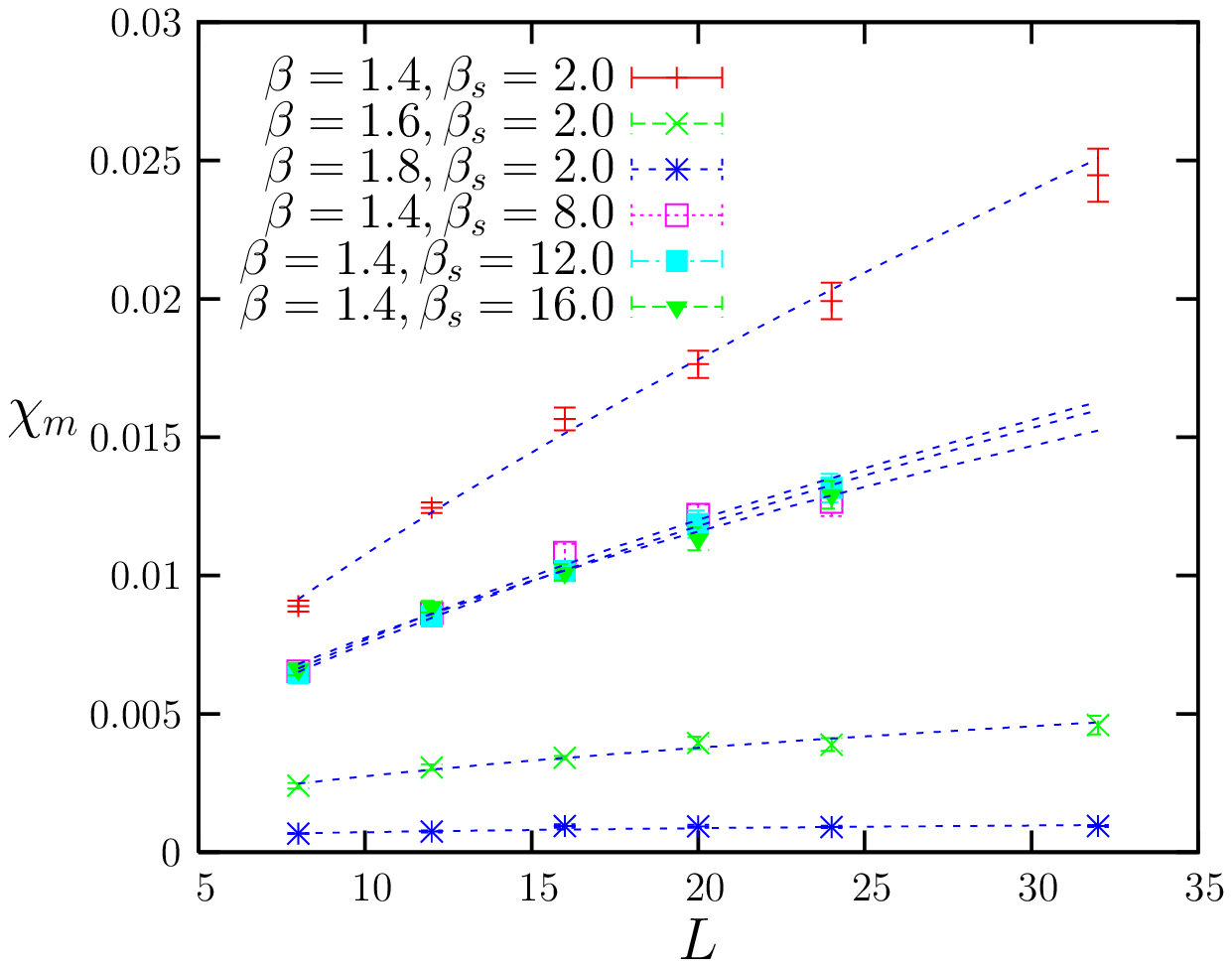}
\\[-1mm]
(b) 
\end{center}
\end{minipage}
\caption{(Color online)
$\chi_m$ vs. $L$ for (a) $N_f=2$, and (b) $N_f=4$. The fitting 
functions (eq.~(\ref{eq:cardy})) are represented by solid lines.
}
\label{fig:chim}  
\end{center}
\end{figure}

\begin{figure}[btp]
    \centering
    \includegraphics[width=9.0cm]{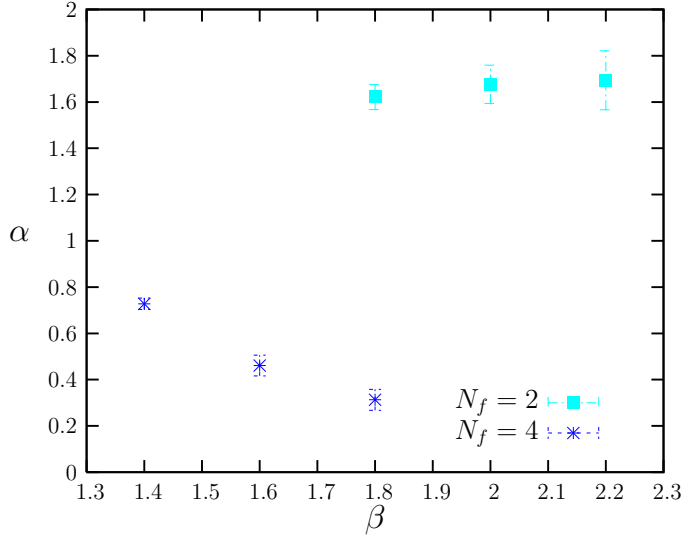}
    \vspace{-3mm}
    \caption{(Color online)
Exponent $\alpha$ vs. $\beta$ extracted from fits with eq.~(\ref{eq:cardy}) for $N_f=2$ and $N_f=4$.
}
\label{gr:alpha}
\end{figure}

Next, we turn our attention to the role of magnetic monopoles. 
As discussed in Sec.~2, the most relevant observable for deciding whether the monopoles are 
in a plasma or a dipolar phase  is 
the monopole susceptibility $\chi_m$. Its short distance contributions $\chi_1$ and $\chi_2$ elucidate 
further the situation. It is instructive to compare data for these observables from  
both the compact and non-compact lattice formulations of QED$_3$. Recent ncQED$_3$ simulations 
\cite{Armour2010} showed that the magnetic charges form tightly bound $m\bar{m}$ dipoles. 
In Figures~\ref{fig:chim} (a) and (b) 
we present the  $\chi_m$ results at weak gauge couplings for $N_f=2$ and $N_f=4$, respectively.  
The data are fitted to a power-law relation 
\begin{equation}
\chi_m = c L^{\alpha} 
\label{eq:cardy}
\end{equation}
and the extracted values of the exponent $\alpha$ are shown in 
Fig.~\ref{gr:alpha}. The clear increase of $\chi_m$ with $L$ in Fig.~\ref{fig:chim} and the non-zero 
values of the exponent $\alpha$ in Fig.~\ref{gr:alpha} imply 
that the magnetic charges are in a plasma phase. 
As expected, the values of $\alpha$ for $N_f=2$ are larger than 
those for $N_f=4$. This can be understood in terms of the renormalization 
group invariant Dirac quantization condition $eg=e_Rg_R=n/2$ ($n$ is an integer 
and the subscript $R$ denotes renormalized charges):
As $N_f$ increases the $e^+e^-$ interaction decreases due to enhanced screening 
from virtual fermion-antifermion pairs, which in turn implies that
the $m \bar{m}$ interaction is antiscreened. In addition the $N_f=4$ values of $\alpha$ appear to 
decrease with $\beta$. It is possible that for large $N_f$ values, the extracted values of $\alpha$ may be
affected by finite size effects. This can be understood as follows: 
The interaction among magnetic dipoles leads
to a screening of the  $m\bar{m}$ interaction \cite{Herbut1,Herbut2}, resulting in a small density of unbound magnetic 
charges. This mechanism implies that the existence of a monopole plasma may be a very long 
distance effect. Therefore, simulations on larger lattices may be required in order to 
extract more accurate values of $\alpha$.
The existence of the monopole plasma depends solely on the gauge field dynamics and it is not expected to 
depend on the four-Fermi coupling provided the latter is weak enough. 
This is supported by the $\beta_s>2.0$ data in Fig.~\ref{fig:chim} (b).  
The data for $\beta_s=8.0, 12.0, 16.0$ from simulations with $L=8,...,24$ almost collapse on a single curve and 
the values of $\alpha$ for the three different $\beta_s$ 
are consistent with $\alpha=0.61(7)$ which is close to the $\beta_s=2.0$ value $\alpha=0.73(3)$. The slightly 
smaller value of $\alpha$ in the $\beta_s \to \infty$ limit could be attributed to the fact that the additional 
four-Fermi term enhances the $e^-e^+$ interaction, implying an enhanced antiscreening effect 
on the $m\bar{m}$ interaction.

\begin{figure}[hbtp]
    \centering
    \includegraphics[width=9.0cm]{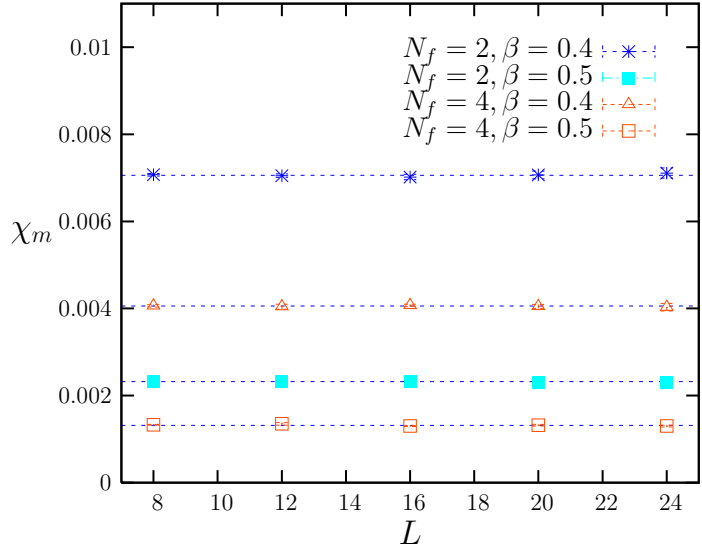}
    \caption{(Color online)
$\chi_m$ vs. $L$ for ncQED$_3$ with $N_f=2, 4$ and $\beta=0.4, 0.5$. The horizontal lines represent fits to the
data.
}
\label{gr:cardy_ncqed3}
\end{figure}

\begin{figure}[tbp]
\begin{center}
\begin{minipage}[c][7.8cm][c]{7.8cm}
\begin{center}
    \includegraphics[width=7.8cm]{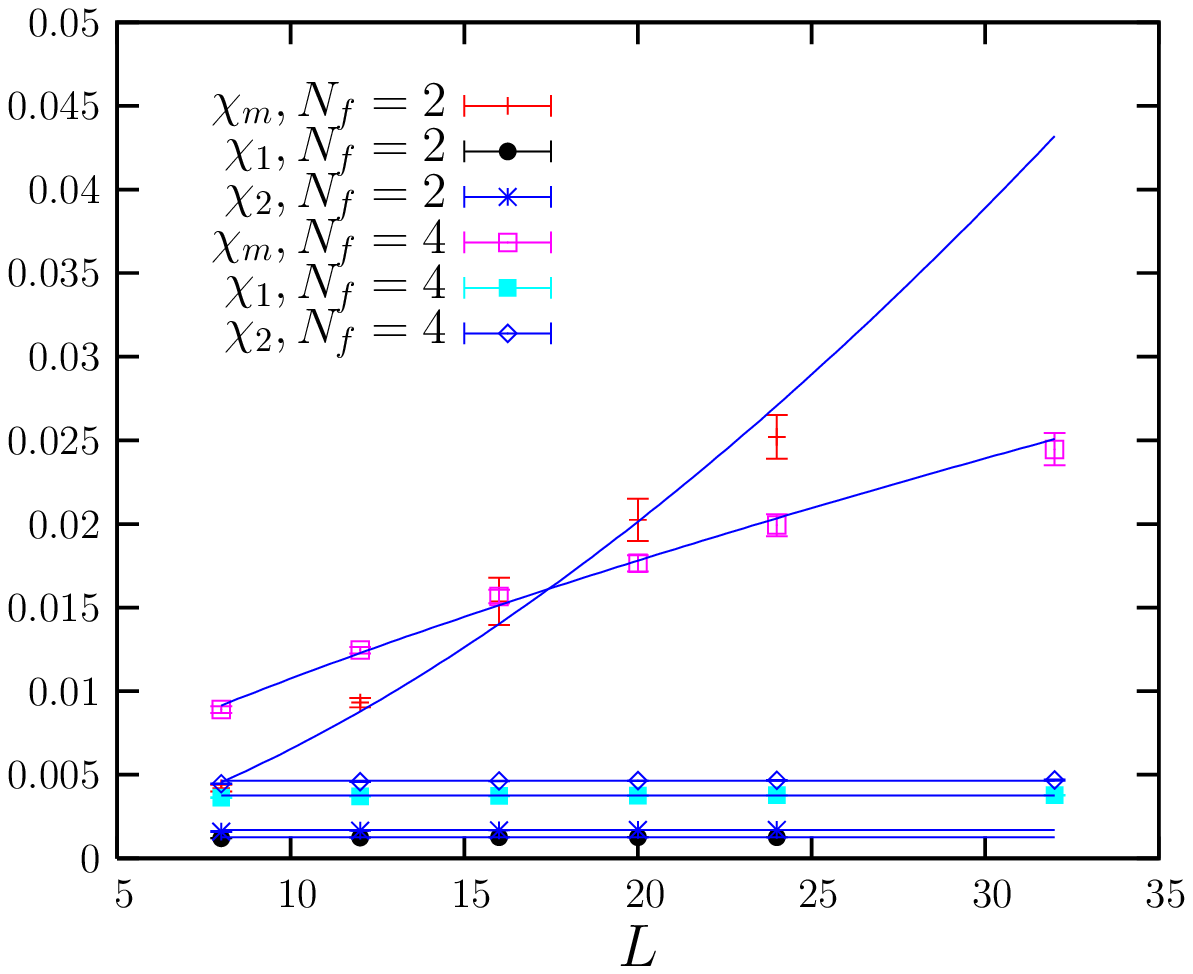}
\\[-1mm]
(a)
\end{center}
\end{minipage}
\begin{minipage}[c][7.8cm][c]{7.8cm}
\begin{center}
    \includegraphics[width=7.8cm]{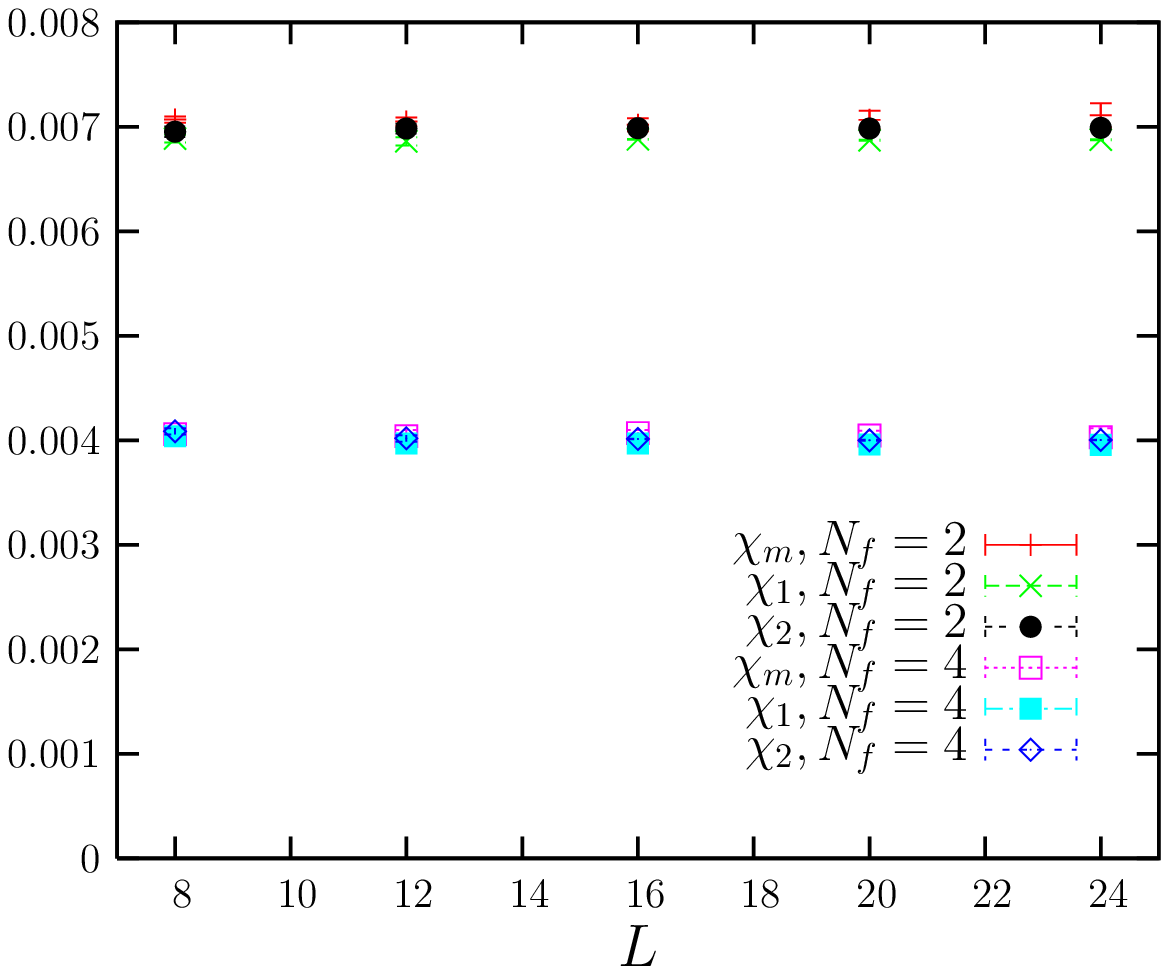}
\\[-1mm]
(b)
\end{center}
\end{minipage}
\caption{(Color online)
$\chi_m$, $\chi_1$, and $\chi_2$ vs. $L$ for (a) cQED$_3$ with $N_f=2, \beta=1.8$ and  $N_f=4, \beta=1.4$
and (b) ncQED$_3$ with $N_f=2,4$ and $\beta=0.4$.
}\label{fig:chi1.2}
\end{center}
\end{figure}

At this point it worth comparing data for $\chi_m$ from both cQED$_3$ and ncQED$_3$. 
In Fig.~\ref{gr:cardy_ncqed3} we present $\chi_m$ versus $L$ for ncQED$_3$ above the chiral transition/crossover 
for $N_f=2, 4$, $\beta_s=2.0$ and $\beta=0.4, 0.5$. The horizontal lines
give excellent fit qualities, implying that the values of $\chi_m$ do
not depend on $L$, because the magnetic charges are in the dipolar
phase.

In Figures~\ref{fig:chi1.2} (a) and (b) we plot $\chi_m$ together with its short distance contributions  
$\chi_1$ 
and $\chi_2$ versus $L$ for cQED$_3$ and ncQED$_3$, respectively. 
For cQED$_3$, the  
$\chi_1$ and $\chi_2$ data fall  
on different horizontal lines below $\chi_m$, because the divergence of $\chi_m$ which is a signature
for a monopole plasma comes from 
long distance contributions. 
In contrast to this,  
the ncQED$_3$ data for $\chi_1$, $\chi_2$ and $\chi_m$
coincide within statistical errors.  This confirms that in ncQED$_3$ 
the contribution to the polarizability comes from 
tightly bound $\bar{m}m$ dipoles which in the continuum ($\beta \to \infty$) limit may disappear by 
collapsing into zero size.

\begin{figure}[tbp]
    \centering 
    \includegraphics[width=8.0cm]{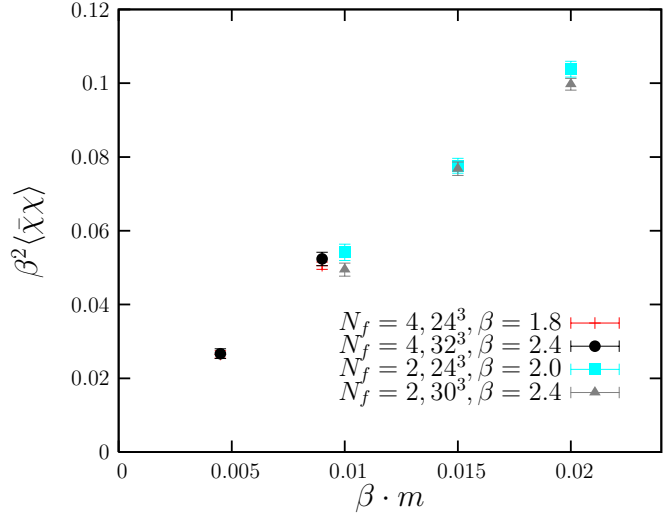}
    \caption{(Color online)
$\beta^2 \langle \bar{\chi}\chi \rangle$ vs $\beta m$ for $N_f=2$, $N_f=4$ at different values of
the coupling $\beta$. The physical volume $(L/\beta)^3$ is constant for each $N_f$ value.
}
\label{fig:continuum}
\end{figure}

Next, we check whether the gauge couplings $\beta=2.0$ and $\beta=1.8$ are in the asymptotic scaling 
regimes for  $N_f=2$ and $N_f=4$, respectively.
Since for an asymptotically-free field theory the ultraviolet behavior is governed 
by the gaussian fixed point at the origin, then the continuum limit of the model lies in the limit 
$\beta \to \infty$, and all physical quantities should be expressible in terms of the scale set by 
the dimensionful coupling $e^2$. To compare simulation results taken at different couplings (lattice spacings),
therefore, it is natural to work in terms of dimensionless variables such as $\beta m$, $L/\beta$, and
$\beta^2 \langle \bar{\chi} \chi \rangle$. As the continuum limit is approached, data taken at different 
$\beta$ should collapse onto a single curve when plotted in dimensionless units. 
To check whether lattice data are characteristic of the continuum limit we
plot the dimensionless chiral condensate $\beta^2 \langle \bar{\chi} \chi
\rangle $ versus the dimensionless fermion bare mass $\beta m$ for $N_f=2$
(with $\beta=2.0$, $L=24$ and $\beta=2.4$, $L=30$) and $N_f=4$ (with $\beta=1.8$
, $L=24$ and $\beta=2.4$, $L=32$ ) in Fig.~\ref{fig:continuum}.
For both $N_f=2$ and $N_f=4$, the values of $\beta^2 \langle \bar{\chi} \chi \rangle$ 
for the two values of $\beta$  agree within 1-2 
standard deviations, implying that lattice discretization effects are small.
\begin{figure}[tb]
\begin{center}
\begin{minipage}[c][7.8cm][c]{7.4cm}
\begin{center}
    \includegraphics[width=7.4cm]{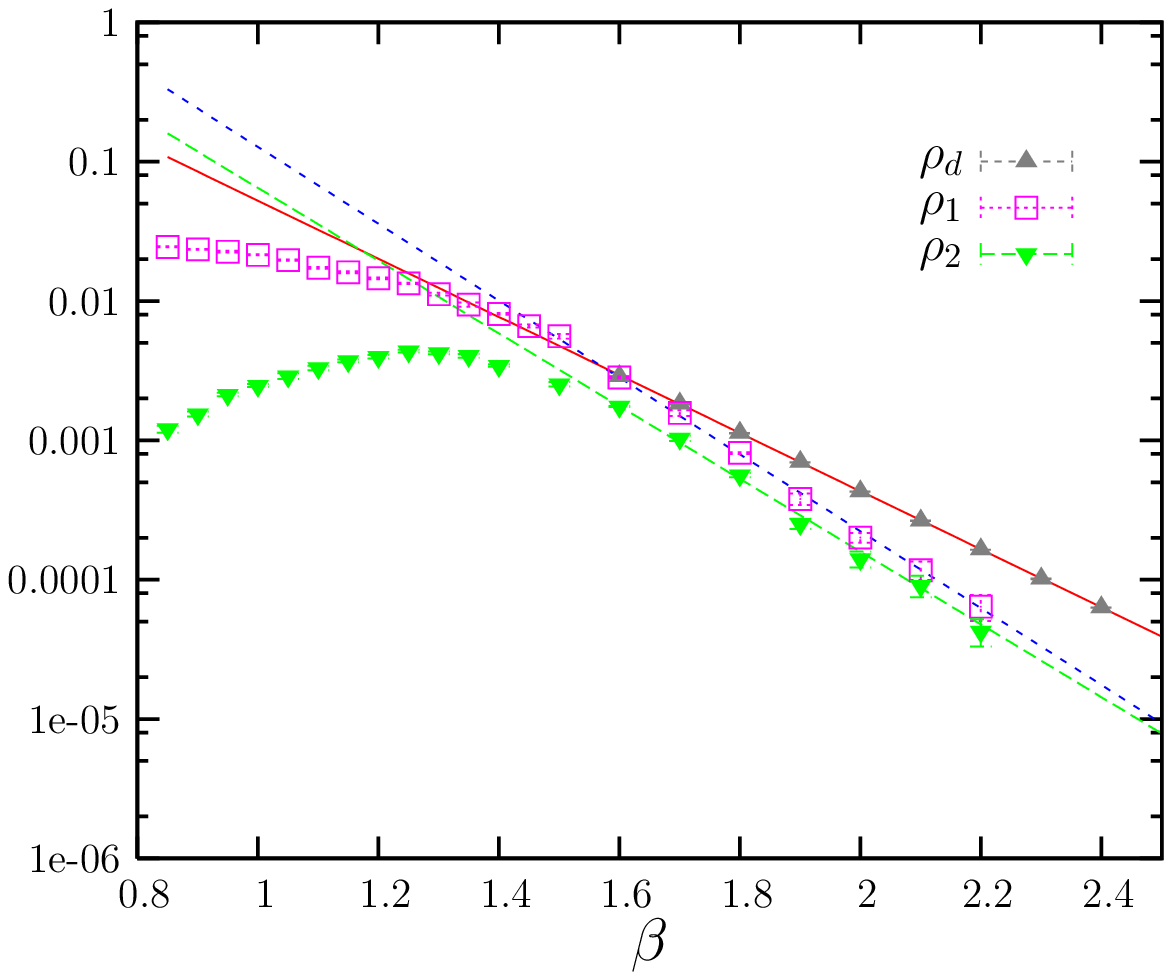}
\\[-1mm]
(a)
\end{center}
\end{minipage}
\begin{minipage}[c][7.8cm][c]{7.4cm}
\begin{center}
    \includegraphics[width=7.4cm]{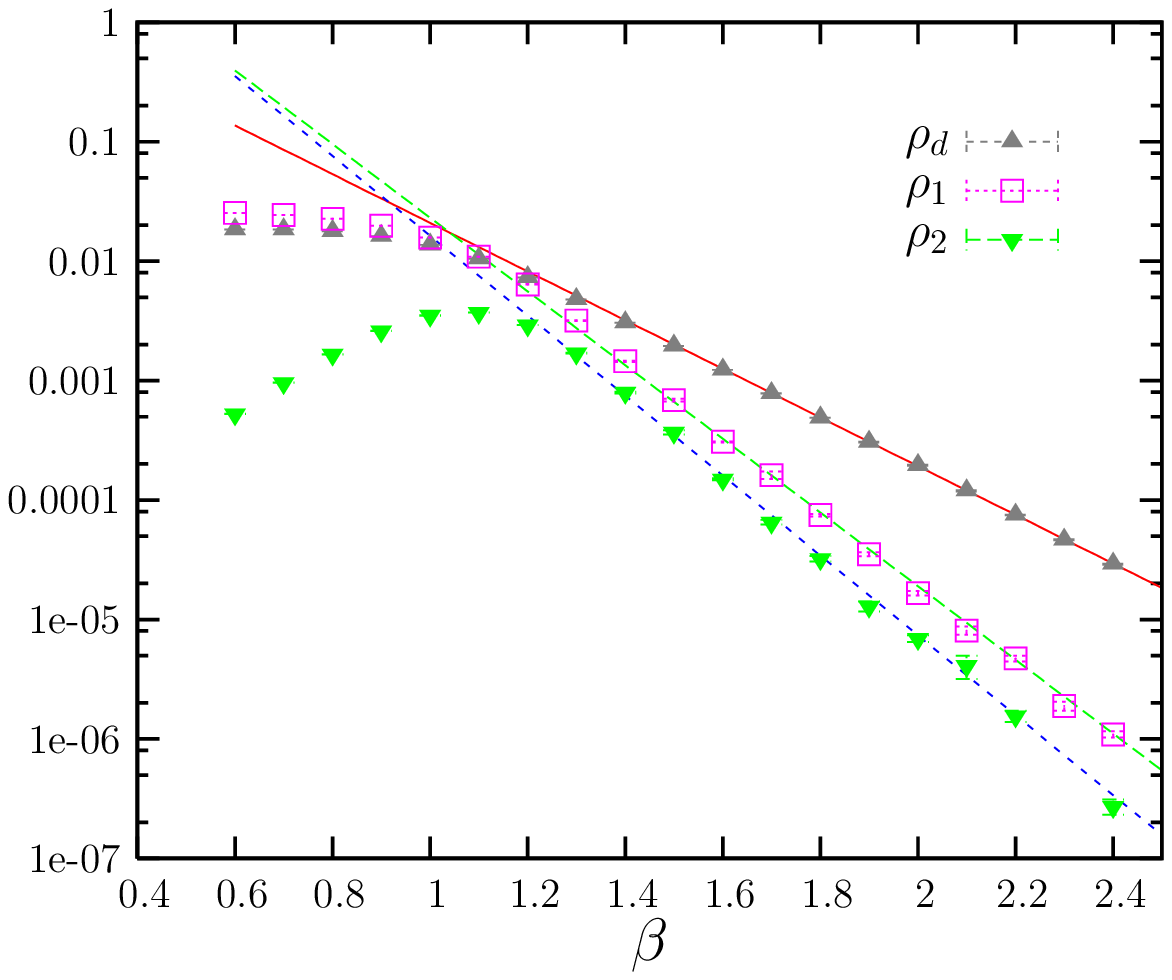}
\\[-1mm]
(b)
\end{center}
\end{minipage}
\caption{(Color online)
$\rho_d$, $\rho_1$, and $\rho_2$ vs. $\beta$ for (a) $N_f=2$, and (b) $N_f=4$.
}
\label{fig:densities}

\end{center}
\end{figure}
\begin{table}[b]
\centering \caption{Values of parameter $\alpha_2$ extracted from fits of $f(\beta)=a_1\exp(-a_2 \beta)$
on $\rho_d$, $\rho_1$ and $\rho_2$ vs. $\beta$ data.}
\medskip
\label{tab:t2}
\setlength{\tabcolsep}{1.5pc}
\begin{tabular}{ccc}
\hline \hline
$N_f$     &    2  &  4      \\
\hline
$a_2 (\rho_d)$ & 4.800(4) &  4.660(3)   \\
$a_2 (\rho_1)$ & 6.4(1)  &  7.13(5)   \\
$a_2 (\rho_2)$ & 6.0(1)  &  7.8(1)   \\

\hline \hline
\end{tabular}
\end{table}

\begin{figure}[btp]
    \centering
    \includegraphics[width=8.0cm]{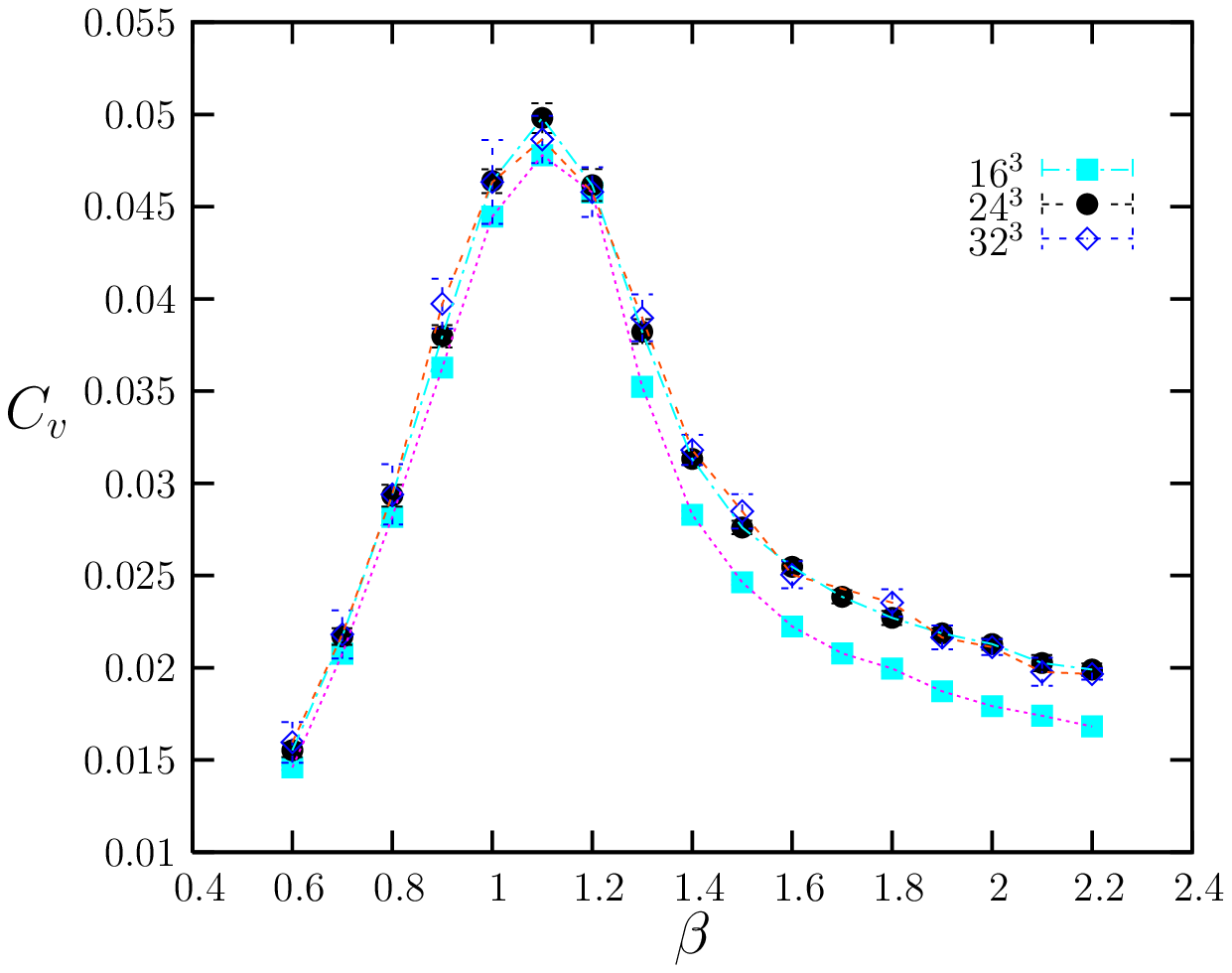}
    \caption{(Color online)
Specific heat $C_v$ vs. $\beta$ for $N_f=4$ and $L=16, 24, 32$.
}
\label{gr:nf=4_specific.heat}
\end{figure}

In order to complete the picture presented in the previous paragraphs  
we also measured various densities: (i) the density $\rho_d$ of ``isolated'' tightly bound dipoles, 
which are $m\bar{m}$ pairs on adjacent cubes, and each charge in the dipole does not share any other 
cube face with with a second opposite charge;
(ii) the density $\rho_1$ of positive magnetic charges that do not share a cube face with an antimonopole; and (iii)
the density $\rho_2$ of   
``isolated'' positive magnetic charges, that not share a cube face, an edge or 
a corner with an opposite charge. 
We fitted the data for $\rho_d$, $\rho_1$ and $\rho_2$ versus $\beta$ to an 
exponential function $f(\beta)=a_1\exp(-a_2\beta)$.
The data and the fitted curves for $N_f=2$ and $N_f=4$ are shown in Figures~\ref{fig:densities} (a) and(b),
respectively. 
 The results
are from simulations on $32^3$ lattices and a comparison with data on $24^3$ lattices showed that
finite volume effects are negligible.
It is clear from the peak values of $\rho_2$ that a 
crossover from a dense monopole plasma phase at strong couplings to a 
dilute monopole gas at weak couplings occurs. The values of the crossover couplings are  
$\beta \approx 1.25$ and $\beta \approx 1.10$ for $N_f=2$ and $N_f=4$, respectively. 
The values of $a_2$ are presented in Table~\ref{tab:t2}. 
The fact that the exponential function accurately fits 
the data implies that there is no phase transition that would result in an abrupt decrease of 
the densities. As expected, the values of the parameter $a_2$ for the ``isolated'' monopoles are a bit larger than 
the respective one in quenched cQED$_3$ where $a_2 \approx 5$ \cite{DeGrand1980}, whereas for  the ``isolated'' 
dipoles the values of $a_2$ are smaller than the quenched value $a_2=6.6$ \cite{DeGrand1980}. 
The crossover from a dense monopole plasma to a dilute monopole gas is also supported  by the behavior of 
the so-called 
specific heat $C_v$ defined in a way analogous to the specific heat in spin models (with the temperature $T$ 
interchanged with the gauge coupling $g^2$) as follows:
\begin{equation}
C_v = -\beta^2 \frac{\partial \langle P \rangle}{\partial \beta}=
\beta^2 \frac{\partial^2 \ln {\cal Z}}{\partial \beta^2}=\beta^2[\langle P^2 \rangle 
-\langle P \rangle^2],
\end{equation}
where ${\cal Z}$ is the lattice partition function and 
$P\equiv \frac{1}{V} \sum_{x,\mu<\nu}(1-\cos\Theta_{\mu\nu x})$ is the pure gauge part of the action 
per unit volume.
In Fig.~\ref{gr:nf=4_specific.heat} we plot $C_v$ versus $\beta$ for $N_f=4$ and $L=16, 24, 32$.
It is clear that $C_v$ develops a lattice size-independent peak at 
$\beta=1.1$ 
The $L$-independent peak of $C_v$ implies that a smooth crossover  
takes place in the gauge field dynamics at $\beta=1.1$, which coincides with the crossover from 
a dense monopole plasma to a dilute monopole gas. 

\section{Conclusions and Outlook}

cQED$_3$ is an interesting field theory due to its similarities with QCD and its close relation 
with QCD-like theories \cite{Unsal2}.
In this paper we presented the first analysis of monopole dynamics in cQED$_3$ with $N_f \leq 4$ 
based on results from  lattice simulations. Fast simulations with massless fermions 
were enabled by adding a weak four-Fermi interaction to the cQED$_3$ action, because the 
vacuum expectation value of the $\sigma$ meson field which appears explicitly in the semi-bosonized 
action acts like a fermion mass and makes the Dirac matrix inversion fast. In addition, in the presence 
of the four-Fermi term the action has an exact $Z_2$ chiral symmetry, which
forbids symmetry breaking lattice discretization counterterms in the free energy. 

The monopole susceptibility (polarizability) diverges with the lattice extent, implying that the monopoles are in 
a plasma phase, which in turn leads to linear confinement of electric charges. Simulations at a 
 single value of $\beta$ for $N_f=4$ showed that the monopole plasma scenario does not depend on the four-Fermi 
coupling $\beta_s$ when $\beta_s$ is sufficiently large. The cQED$_3$ results for the monopole susceptibility 
were contrasted with ncQED$_3$ 
data where $\chi_m$ is independent of $L$ and its major contribution comes from charges on adjacent lattice cubes, 
implying that the single lattice spacing size dipoles at finite $\beta$ may collapse to zero size in the 
continuum limit. In addition, the behavior of the
effective chiral order parameter $\langle | \bar{\chi} \chi | \rangle$ for $N_f=4$ implies that a crossover 
instead of a transition takes place at strong couplings, which could be an outcome of a $V(r) \sim r$ 
electric potential. Also, the behavior of the density of ``isolated'' 
monopoles favors a scenario of a crossover from a dense plasma of monopoles 
to a dilute monopole gas at weak couplings. This scenario is supported by the $L$-independent peak of the 
so-called specific heat. 
Our results imply that for $N_f \leq 4$ the continuum limits of cQED$_3$ and ncQED$_3$ are different, with linear 
charge confinement for the former and logarithmic confinement for the later provided that $N_f=4$ is below the 
ncQED$_3$ $N_f^{\rm crit}$. 

Our conclusions are supported by the results of~\cite{Hands:2006dh},
where it was shown that an isolated magnetic charge has an infinite
free energy, both for the dynamical and the quenched system. Following similar lines,
we plan to test the response of the system to
the insertion of a static dipole. A vanishing free energy gap would
confirm our results that monopoles are in a neutral plasma phase.

The existence of a monopole plasma phase in cQED$_3$ has implications in strongly correlated electron
systems. More specifically $U(1)$ spin liquids (staggered-flux spin liquid corresponds to $N_f=2$ 
and $\pi$-flux spin liquid to $N_f=4$) in two spatial dimensions which are believed to describe the 
underdoped Mott insulator regime 
in cuprate superconductors may be unstable to spinon confinement. It should be noted, however, 
that the anisotropic interactions of these condensed matter systems have been neglected in our model. Recent 
analytical \cite{Concha} and numerical \cite{Thomas} results of ncQED$_3$ 
with  Fermi and gap anisotropies showed that the velocity anisotropy 
is relevant  in the RG sense and its increase leads to a decrease of $N_f^{\rm crit}$.  

We are currently expanding the cQED$3$ simulations to larger $N_f$ values. The  plan is to search for a 
putative conformally invariant fixed point at $N_f^{\rm crit}$ where a phase transition from a linearly 
confining phase to conformal deconfined phase may 
take place. In the condensed matter language this critical point would correspond to a deconfined quantum 
critical point \cite{Sachdev}, where a phase transition is expected to occur between a 
phase with N\`{e}el aniferromagnetic order (at small $N_f$)
and a paramagnetic critical spin liquid phase (at large $N_f$). 
In addition, as emphasized in \cite{Lee2006} $(2+1)d$ two-color QCD
may provide a more appropriate description of algebraic spin liquids than cQED$_3$ which only includes Gaussian 
fluctuations about the mean field and suffers from various limitations in the underdoped regime. 
The study of the phase diagram of $(2+1)d$ two-color QCD is another non-perturbative 
problem that requires lattice simulations for reaching definitive answers.

\section*{Acknowledgements}
The authors wish to thank the Diamond Light Source for kindly allowing
them to use extensive computing resources and the Oxford University
Supercomputing Centre for computing time on the ORAC supercomputer.
The work of B.L. is supported by the Royal Society through the University
Research Fellowship scheme. B.L. also acknowledges partial financial
support by STFC under contract ST/G000506/1.

\end{document}